# Z-checker: A framework for assessing lossy compression of scientific data

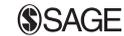

Dingwen Tao[1], Sheng Di[2], Hanqi Guo[2], Zizhong Chen[1,3]
and Franck Cappello[2,4]

## Abstract
Because of the vast volume of data being produced by today's scientific simulations and experiments, lossy data compressor allowing user-controlled loss of accuracy during the compression is a relevant solution for significantly reducing the data size. However, lossy compressor developers and users are missing a tool to explore the features of scientific data sets and understand the data alteration after compression in a systematic and reliable way. To address this gap, we have designed and implemented a generic framework called Z-checker. On the one hand, Z-checker combines a battery of data analysis components for data compression. On the other hand, Z-checker is implemented as an open-source community tool to which users and developers can contribute and add new analysis components based on their additional analysis demands. In this article, we present a survey of existing lossy compressors. Then, we describe the design framework of Z-checker, in which we integrated evaluation metrics proposed in prior work as well as other analysis tools. Specifically, for lossy compressor developers, Z-checker can be used to characterize critical properties (such as entropy, distribution, power spectrum, principal component analysis, and autocorrelation) of any data set to improve compression strategies. For lossy compression users, Z-checker can detect the compression quality (compression ratio and bit rate) and provide various global distortion analysis comparing the original data with the decompressed data (peak signal-to-noise ratio, normalized mean squared error, rate–distortion, rate-compression error, spectral, distribution, and derivatives) and statistical analysis of the compression error (maximum, minimum, and average error; autocorrelation; and distribution of errors). Z-checker can perform the analysis with either coarse granularity (throughout the whole data set) or fine granularity (by user-defined blocks), such that the users and developers can select the best fit, adaptive compressors for different parts of the data set. Z-checker features a visualization interface displaying all analysis results in addition to some basic views of the data sets such as time series. To the best of our knowledge, Z-checker is the first tool designed to assess lossy compression comprehensively for scientific data sets.

## 1. Introduction

One of the most challenging issues in performing scientific simulations or running large-scale parallel applications today is the vast amount of data to store in disks, to transmit on networks, or to process in postanalysis. The Hardware/Hybrid Accelerated Cosmology Code (HACC), for example, can generate 20 PB of data for a single one-trillion-particle simulation (Habib et al., 2016); yet a system such as the Mira supercomputer at the Argonne Leadership Computing Facility has only 26 PB of file system storage, and a single user cannot request 75% of the total storage capacity for a simulation. In climate research, as indicated by Gleckler et al. (2016), nearly 2.5 PB of data were produced by the Community Earth System Model (CESM) for the Coupled Model Intercomparison Project (CMIP), which further introduced 170 TB of postprocessing data submitted to the Earth System Grid (Bernholdt et al., 2005). Estimates of the raw data requirements for the CMIP6 project exceed 10 PB (Baker et al., 2014). Particle accelerator–based research also deals with

[1] Department of Computer Science and Engineering, University of California, Riverside, CA, USA
[2] Division of Computer Science and Mathematics, Argonne National Laboratory, Lemont, IL, USA
[3] Beijing University of Technology, Beijing, China
[4] Parallel Computing Institute, University of Illinois Urbana–Champaign, Champaign, IL, USA

**Corresponding author:**
Franck Cappello, Mathematics and Computer Science Division, Argonne National Laboratory, 9700 Cass Avenue, Lemont, IL 60439, USA.
Email: cappello@mcs.anl.gov

a large volume of data during experiment and postanalysis. The Advanced Photon Source (APS) Upgrade (Austin, 2016) is the next-generation APS project at Argonne; one of its targets is mouse brain images of 1.5 cm$^3$. As indicated by APS-based researchers at Argonne, the instruments can produce 150 TB of data per specimen, with a total of at least 100 specimens for storage. Also involved are 35 TB of postprocessing data. Obviously, effective data compression is significant for the success of today's scientific research.

Lossy compressors are commonly considered the relevant solution for significantly shrinking the data size for scientific research, while still carefully controlling the loss of data accuracy based on user demand. However, scientific data sets are often composed of floating-point data arrays, which lossless compressors such as GZIP (Deutsch, 1996) cannot compress effectively (Baker et al., 2016). Arguably, many floating-point data-based lossy compressors have been proposed, some of which are considered relevant solution for climate simulations by guaranteeing the validity of data after decompression (Baker et al., 2016; Sasaki et al., 2015). However, many other types of scientific research issues need lossy compression, and the impact of lossy compressors on their results is still unclear. Moreover, users want to be able to select the best-fit compressors for their specific needs.

Clearly needed, then, an easy-to-use, generic, comprehensive assessment tool for users to understand the effectiveness of lossy compressors and their impact on their scientific results. The key challenges in designing such a tool are the masses of evaluation metrics to deal with and the various design principles across lossy compressors. On the one hand, scientific researchers often have different targets with different requirements on the accuracy of data. For instance, graph-processing researchers focus mainly on peak signal-to-noise ratio (PSNR) (Taubman and Marcellin, 2012), while climate researchers may also be concerned with other metrics such as the correlation coefficient of the original data and decompressed data (Baker et al., 2014). Cosmology researchers, as indicated by the HACC development team at Argonne, are concerned with the distribution of the compression errors and their autocorrelations. On the other hand, lossy compressors may have largely different designs, such that their compression results can be different even with the same data sets. In particular, we observe that SZ-1.4 (Tao et al., 2017) has a uniform distribution of the compression errors but ZFP has a normal distribution and that ZFP has higher autocorrelation of the compression errors than does SZ-1.4 (see the experimental evaluation section for details). Thus, it is a nontrivial task to design a generic assessment tool for assessing multiple lossy compressors in a comprehensive and fair way.

In this work, we present a flexible, generic, open-source lossy compression assessment tool called Z-checker, which can be used to evaluate the compression quality of various lossy compressors. Z-checker is developed in the context of the Exascale Computing Project (ECP) to respond Exascale application needs. These applications are currently not using lossy compression. Also, since there is no tool to provide a comprehensive assessment of compression quality (such as compression error), Z-check is designed to address this gap. Our key contributions are as follows:

- We design a flexible assessment framework, called Z-checker, that integrates a series of assessment modules and visualization tools. In particular, this framework allows users to assess the lossy compression qualities in multiple ways (such as reading standalone data files, library calls in the compression codes, and external calls of compression executables from Z-checker).

- We integrate in Z-checker assessment algorithms and functions that are as comprehensive as possible. (1) Z-checker can be used to characterize critical properties (such as entropy, distribution, power spectrum, principal component analysis, and autocorrelation) of any data set, such that the difficulty of data compression can be presented clearly in the granularity of data blocks. (2) Not only Z-checker is able to check the compression quality (compression ratio (CR) and bit rate (BR)), but it also provides various global distortion analysis comparing the original data with the decompressed data (PSNR, normalized mean squared error (MSE), rate–distortion, rate-compression error, spectral, distribution, and derivatives) and statistical analysis of the compression error (maximum/minimum/average error, autocorrelation, and distribution of errors). (3) Z-checker can also assess the impact of the lossy decompressed data on some common transform functions, such as discrete Fourier transform (DFT) and discrete wavelet transform (DWT). (4) Z-checker also provides two ways to visualize the data and compression results on demand. Specifically, Z-checker may help generate data figures by static scripts or by an interactive system.

- We implemented the Z-checker software and will release it as an open-source community tool, under a BSD license. To the best of our knowledge, Z-checker is the first tool designed to comprehensively assess compression results for scientific data sets across multiple lossy compressors.

The rest of the article is organized as follows. First, we discuss the research background and design motivation of this tool. Next, we present an overview of Z-checker, including the design architecture, the most commonly used module "user interface module," and the analysis kernels. Then, we describe the analysis and assessment functions in more detail, and we present the evaluation results of our proposed framework. We discuss the related work and conclude the article with a summary.

**Table 1.** Existing lossy compressors.

| Compressor | VQ | Transform | Prediction | BA |
|---|---|---|---|---|
| NUMARCK | ✓ | | | |
| ISABELA | | | ✓ | |
| ZFP | | ✓ | | ✓ |
| SZ | | | ✓ | ✓ |
| FPZIP | | | | ✓ |
| SSEM | ✓ | ✓ | | |

VQ: vector quantization.

**Table 2.** Four use cases of Z-checker.

| | Off-line processing | Online processing |
|---|---|---|
| Static display | Load data once for all, and view results by generating local image files | Load data based on dynamic requests, and view results by local image files |
| Interactive display | Load data once for all, and view results on demand via a web page interactively | Load data dynamically, and view results on demand by a web page interactively |

## 2. Research background and design motivation

In this section, we present background information intended to help readers better understand today's scientific simulation data sets and existing lossy compressors.

A vast majority of scientific simulation data (especially the postanalysis data produced by scientists) is composed of floating-point data arrays. Thus, floating-point data compression is particularly needed for data reduction when the scientific data size is too huge to be dealt with in a reasonable time or to be stored in disks or parallel file systems (PFS).

Many lossy data compressors (as listed in Table 1) have been designed to significantly shrink the scientific data stored in the form of floating-point arrays. State-of-the-art lossy compressors often combine multiple strategies, such as vector quantization (VQ), orthogonal transform, prediction, and analysis of floating-point binary representation (BA). NUMARCK (Chen et al., 2014), for example, approximates the differences between snapshots by VQ. ISABELA (Lakshminarasimhan et al., 2013) converts the multidimensional data to a *sorted* data series and then performs B-spline interpolation. ZFP (Lindstrom, 2014) involves more complicated techniques such as fixed-point integer conversion, block transform, and bit-plane encoding. FPZIP (Lindstrom and Isenburg, 2006) adopts predictive coding and also ignores insignificant bit planes in the mantissa based on the analysis of the IEEE 754 (Committee et al., 2008) binary representation. SSEM (Sasaki et al., 2015) splits data into a high-frequency part and low-frequency part by wavelet transform and then uses VQ and GZIP. SZ is an error-bounded lossy compressor proposed by Di and Cappello (2016) and Tao et al. (2017); it comprises four compression steps: (1) perform multidimensional prediction for each data point by its neighbor points' values, (2) encode the prediction error by a uniform scalar quantization method, (3) perform the binary analysis for unpredictable data, and (4) perform lossless compression such as Huffman encoding (Huffman et al., 1952) and the LZ77 algorithm (Ziv and Lempel, 1977).

In general, the listed lossy compression tools can be used in two ways for reducing the data size, and our assessment tool supports both ways. On the one hand, they generally provide straightforward executables to perform the compression on original data files stored in disks or do the decompression on the compressed data files directly. On the other hand, they provide easy-to-use library interfaces such that scientific simulation programs can call the corresponding functions to do the in situ compression at runtime. This latter approach is especially useful for improving the I/O performance for high-performance computing (HPC) applications because the data have been divided by the users for parallel computing and can be compressed by each process or rank in parallel.

## 3. Overview of Z-checker framework

In this article, we present Z-checker, a novel framework with three important features: (1) Z-checker can be used to explore the properties of original data sets for the purpose of data analytics or improvement of lossy compression algorithms. (2) Z-checker is integrated with a rich set of evaluation algorithms and assessment functions for selecting best-fit lossy compressors for specific data sets. (3) Z-checker features both static data visualization scripts and an interactive visualization system, which can generate visual results on demand. This interactive mode allows compression algorithm developers and users to compute dynamically analysis on user-selected portions of the data set.

Z-checker is designed to support two processing modes: online and off-line, and two display modes: static and interactive. The four modes correspond to four different use cases. The off-line static mode is useful for generating a report on the original data set properties and the compression error, presenting multiple analysis views. The off-line interactive mode is useful for exploring the properties of the original data sets and the compression error. It allows users to dig into some region of interest and reveal local properties that are not present or visible on the whole data set. The online static display mode corresponds to in situ data compression and may generate and update the original data set compression analysis as long as the data are presented to Z-checker. The online interactive mode allows users to monitor the compression error while the data are produced. Users can zoom in on regions of interest and assess the quality of the data compression. Table 2 summarizes the four use cases of Z-checker, considering the two processing modes and two visualization modes.

The design architecture of Z-checker is presented in Figure 1, which involves three critical parts: *user interface*, *processing module*, and *data module*.

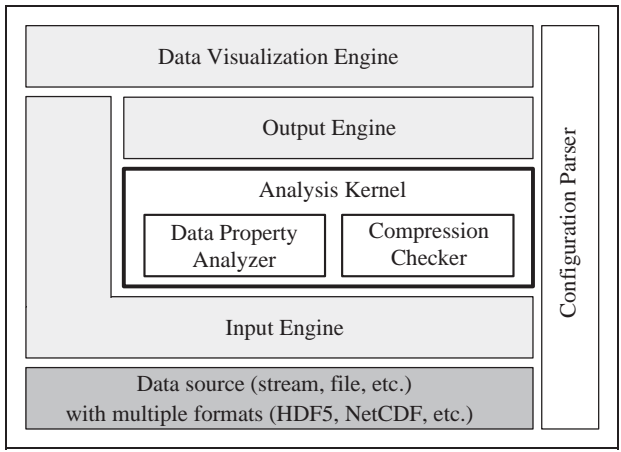

**Figure 1.** Design architecture of Z-checker.

- *User interface* includes three key engines—input engine, output engine, and data visualization engine—as shown in the light-gray rectangles in Figure 1. They are in charge of reading the floating-point data stream (either original data or compressed bytes), dumping the analyzed data to disks/PFS, and plotting data figures for visualizing analysis results. The data visualization engine also provides the interactive mode through a web browser interface (details are described later).
- *Processing module*, in the whole framework, is the core module, which includes the analysis kernel and configuration parser. The former is responsible for performing the critical analysis, and the latter is in charge of parsing the user's analysis requirements (such as specifying input file path, specifying the compression command or executable, and customizing the analysis metrics on demand). Specifically, the analysis kernel is composed of two critical submodules, namely, the data property analyzer and compression checker, which are responsible for exploring data properties based on the original data sets and analyzing the compression results with specified lossy compressors (discussed later in more detail).
- *Data source module* (shown as the dark-gray box in the figure) is the bottom layer in the whole framework and represents the data source (such as data stream produced by scientific applications at runtime or the data files stored in the disks).

In what follows, we describe the user interface module and analysis kernel, respectively.

## 3.1. User interface module of Z-checker

In the user interface module, the input engine is used to retrieve the data stream and convert it to have the correct format if necessary. Depending on different ways of gleaning data, Z-checker allows users to perform the data property analysis and compression checking in three ways[1]: (1)

reading data files stored in disks and performing off-line analysis, (2) getting the data stream in the user's scientific simulation codes and performing online analysis, and (3) calling external compression commands or executables and checking the compression results based on system libraries or commands such as *time()*. Note that the accuracy of *time()* is generally good enough because compression of large data sets can take tens of minutes. However, other more accurate timing commands can be added by the users if needed. Moreover, the input engine needs to do some preprocessing steps for the analysis of the data or compression results. For instance, it needs to check the endianness of the current system and guarantee the correct parsing order of the data (either big-endian type or little-endian type, which is specified by users in a configuration file). The data may also be stored in different file formats such as HDF5 and NetCDF, such that they should be parsed by the input engine before the analysis. Note that currently the input engine assumes that the original data set and the decompressed data set have the same size (the same number of data points). Some data reduction techniques, such as decimation, reduce the size of the data set by removing data points. The future version of Z-checker will update the input module to deal with decimated data sets.

The output engine in the user interface module is used to construct the output data (such as properties of original data and compression results under various compressors) based on the analysis results produced by the analysis kernel. Specifically, it converts the analysis results to be consistent with the format required by the visualization toolkits.

The visualization engine contains two parts: a static visualization tool and an interactive visualization system. The former generates a battery of Gnuplot (Janert, 2009) scripts, which can be used to plot the data conveniently in various formats (such as eps, jpg, and png) by the Gnuplot library. The latter is an efficient web system, providing an interactive web page for users to submit and specify their analysis demands, compression metrics, and comparison requirements and to visualize the results that are generated in the backend servers. In what follows, we first describe the design of our static visualization tool and then discuss how the interactive visualization system is performed based on the analysis results generated by the output engine.

### 3.1.1. Design and implementation of static visualization.
We integrate in our static visualization tool a set of Gnuplot templates, including various types of figures such as lines, linespoints, histograms (bar chart), and fillsteps, which are used to plot different analysis results for users. The histogram figure, for example, is used to present the evaluation metrics such as CR, compression time, and rate across different variables and compressors. The fillsteps figure is used to show some properties of the original data and the distribution of compression errors. The linespoints figure is used to present the comparison results with multiple compression requirements and compressors such as the rate–distortion of compression data.

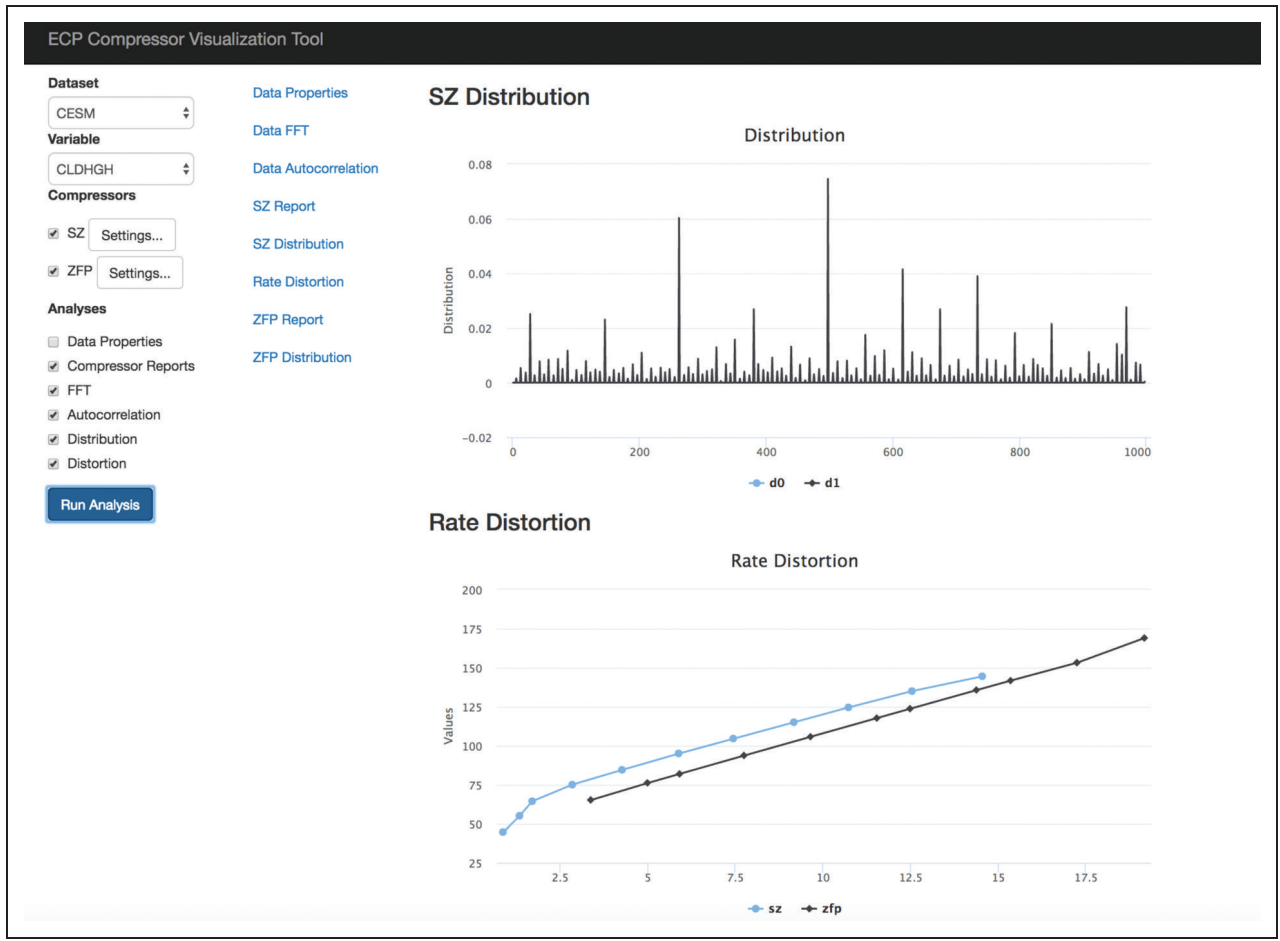

**Figure 2.** Screenshot of Z-checker's interactive visualization system.

*3.1.2. Design and implementation of interactive visualization system.* The interactive visualization system displays the data figures based on the user's requirements specified via an interactive web page. Figure 2 shows a screen shot of the interactive visualization system, which, as an example, displays the rate–distortion results of SZ versus ZFP based on the variable CLHHGH from CESM climate simulation. More analysis results are shown in the evaluation section.

The implementation of the interactive visualization system is based on a client–server model. Specifically, it receives the user's submitted request, executes the corresponding checking commands or analysis work in the backend servers, and plots on the web page the data generated by the output engine.

The server hosts the static web contents and responds to the requests from the clients. The implementation is based on node.js,[2] which is a JavaScript runtime. Once a query is received by the server, the server translates the query into a Z-checker configuration file, calls the Z-checker command line with the configuration file, and then parses and sends the analysis results in JavaScript Object Notation format that can be directly parsed and visualized by the web clients. In addition, we cache the analysis results for the same requests, in order to accelerate the interactive queries. Note that while the server of the visualization engine is using

JavaScript, the analysis kernel is written in C for high performance.

The client side has two components—the query form and the visualization results. The query form allows users to choose the data set, variable, and different compressors for analysis. Users can also choose different analyses, such as error distribution, autocorrelation, and distortion. Once the form is submitted, the web client asynchronously waits for the results from the server. The results are further plotted and visualized with Highcharts.js[3] and d3.js[4] libraries.

## 3.2. Analysis kernel of Z-checker

*3.2.1. Data property analyzer for analyzing scientific data.* One must understand the properties of scientific data sets in order to develop an efficient lossy compressor. To this end, the Z-checker framework is designed to analyze and explore the properties of the original data sets with respect to the compression methods or other analysis metrics demanded by users. The key property analysis functions of our framework for the scientific data sets are listed as follows. In this section, we list the metrics and mainly describe their significance and usage in the framework. We define and discuss them in more detail in the next section.

1. Basic properties, including minimum/maximum/average value and value range: Such basic properties are fundamental in that they will also be utilized in computing other metrics and functions in the framework.

2. Data distribution: This function gives more in-depth understanding of the value range of the data and its density rather than the basic properties. It will also be used in computing other metrics such as the distribution of compression errors.

3. Entropy: This metric represents the Shannon entropy of the data. Splitting the whole data set into multiple data blocks and computing the entropy value for each block may identify the hard-to-compress regions of the data set.

4. Smoothness: The smoothness estimation of the original data can help lossy compression developers decide whether to adopt smoothness techniques, such as sorting and wavelet transform, to smoothen the data sets before applying other compression techniques.

5. Power spectrum: According to our communication with scientific researchers, some are interested in studying the power spectrum of the data, which may be easier for them to understand the nature of the data. This function can also help identify the periodicity of the data sequence for improving the compression quality based on repeated patterns in the sequence of the data.

6. PCA: PCA, such as singular value decomposition, is a fundamental technique adopted by some compressors (Austin et al., 2015), which can significantly improve the CR by leveraging the similarity of different variables and snapshots.

7. Autocorrelation of original data sets: This metric with multiple delay lags can represent the seasonality of the data and explore possible patterns accordingly. It also shows the presence (or not) of autocorrelation in the original data set. This is important to assess the applicability of compressors because some compressors add significantly autocorrelation in the compressed data set.

### 3.2.2. Compression checker for assessing lossy compressors.

Our proposed framework has plentiful algorithms and functions for assessing lossy compressors. The compression methods focus mainly on scientific data sets produced or processed by HPC applications or instruments. Specifically, the HPC applications may generate multiple snapshots containing many variables, each of which has a specific data type such as a multidimensional floating-point array. Our framework is designed to assess lossy compressors based on the following metrics and functions:

1. Pointwise compression error between original and reconstructed data sets, for example, absolute error and value range–based relative error.[5] This metric is widely used by lossy compressors such as SZ (Di and Cappello, 2016; Tao et al., 2017) and ZFP (Lindstrom, 2014), in order to bound the compression errors for users.

2. Statistical compression error between original and reconstructed data sets, such as root MSE (RMSE), normalized RMSE (NRMSE), and PSNR: These metrics are widely used by many scientific researchers (Taubman and Marcellin, 2012) to evaluate the distortion of the data after lossy compression.

3. Distribution of compression errors: Some scientific researchers require the compression errors to follow Gaussian white noise distribution, because a lot of analyses are based on this assumption.

4. CR or BR: This is one of the most important factors to evaluate the compression performance. The higher the CR, the lower the BR, for user-acceptable distortions (as we will see later, there are many relevant metrics to assess the distortion), and then the better the compressor is in general.

5. Rate–distortion based on statistical compression error and BR: Rate–distortion is generally a significant metric with respect to the overall compression quality. Rate refers as the mean number of bits used to represent a data point's value after the compression, while distortion refers to the overall deviation of the data after the compression and is generally assessed via PSNR.

6. Compression and decompression speed (i.e. compression rate): Compression/decompression speed is particularly important for the in situ compression, in which the data compression is executed by each parallel process before dumping the data to the disks. Not only do the users hope to get a high CR in order to save I/O time during the execution, but the compression also has to suffer from limited compression/decompression time such that the overall execution performance can be maximized.

7. Pearson correlation between original and reconstructed data sets: This metric is of particular interest to some scientific users for assessing the deviation of the compressed data set from the original one (Baker et al., 2014).

8. Autocorrelation of compression errors: This metric is important for assessing the degree of autocorrelation (if any) that the lossy compressors add to the original data set.

9. Comparison of derivatives on original and reconstructed data sets: This evaluation is performed by some researchers (Lindstrom, 2014) who are concerned that computations of derived data may amplify any compression errors, resulting in visual artifacts or numerical inaccuracies that were not readily apparent in the original data.

10. Comparison of data transforms on original and reconstructed data sets, such as DFT and DWT:

Minimizing the distortion of such transforms is required by some scientists such as cosmology researchers.

To query the compression results efficiently, we organize all the information by a hashtable, in which the key is composed of the compressor's name and a certain compression error control (such as compression error bound). The hashtable contains the corresponding compression results, such as distribution of compression errors, CR, and compression speed.

# 4. Description of assessment and analysis functions in Z-checker framework

In this section, we describe the assessment algorithms and functions in detail. We first describe the properties that our framework can analyze for scientific data sets. We then describe the compression-based metrics that our framework computes for assessing lossy compressors.

For simplicity, we define some notation as follows. We denote the original multidimensional floating-point data set by $X = \{x_1, x_2, ..., x_N\}$, where each $x_i$ is a floating-point scalar, and we denote the reconstructed data set by $\tilde{X} = \{\tilde{x}_1, \tilde{x}_2, ..., \tilde{x}_N\}$, which is recovered by the decompression process. We also denote the range of $X$ by $R_X$, that is, $R_X = x_{\max} - x_{\min}$.

## 4.1. Properties in data property analyzer of Z-checker

In this subsection, we describe the properties that our framework can analyze for scientific data.

1. To get a general understanding of a scientific data set, our framework analyzes the basic properties of the data set, such as minimum value, maximum value, value range, and average value. Such fundamental metrics are also used in computing other metrics and functions in the framework. For example, the value range is used for calculating the value range–based relative error for assessing lossy compressors, and the average value is used for the autocorrelation analysis.
2. To get the distribution of a data set, our framework divides the value range into multiple equal-length bins and counts the number of data values in each bin. Based on these numbers, the visualizer in our framework plots an approximate figure of the data distribution. The accuracy of the approximate figure depends on the user-set number of bins, which is set to 1000 as default. More specifically, the computation generates the probability density function (PDF) and the cumulative distribution function.
3. For lossless encoding or compression, entropy (more specifically, Shannon entropy) provides an absolute limit on the best possible average length of an information source. Thus, we adopt this metric to evaluate the compressibility of scientific data sets. Generally, the entropy $H$ of a discrete random variable $X$ and probability mass function $P(X)$ is

$$H = E[-\log_2(P(X))] \tag{1}$$

where $E[]$ is the expected value operator.

Specifically, considering lossy compression, we can truncate the original floating-point data according to user-set error bound (accuracy) and calculate the probability mass function. Let the original floating-point data set be $X = \{x_1, x_2, ..., x_N\}$; let the truncated data set be $X^{\text{trun}} = \{x_1^{\text{trun}}, x_2^{\text{trun}}, ..., x_N^{\text{trun}}\}$, where $x_i^{\text{trun}} = eb_{\text{abs}} \lfloor x_i / eb_{\text{abs}} \rfloor$ and $eb_{\text{abs}}$ is the user-set absolute error bound. Then, we merge the same value in $X^{\text{trun}}$, shrink the data set $X^{\text{trun}}$ to $X_{\text{shrk}}^{\text{trun}} = \{x_1^{\text{trun}}, x_2^{\text{trun}}, ..., x_n^{\text{trun}}\}$, and count the probability $P(x_i^{\text{trun}})$ of the value $x_i^{\text{trun}}$. The entropy of the data set $X$ in terms of absolute error bound $eb_{\text{abs}}$ is calculated as follows

$$H(X, eb_{\text{abs}}) = -\sum_{i=1}^{n} P(x_i^{\text{trun}}) \log(P(x_i^{\text{trun}})) \tag{2}$$

Unlike Shannon entropy on lossless compression representing the limit of the average length of lossless encoding for an input source, this value only reflects the compressibility under a certain error bound. Generally speaking, the lower the value of $H(X, eb_{\text{abs}})$ is, the higher compressibility the information source $X$ can achieve, with respect to the error bound $eb_{\text{abs}}$.

In order to evaluate the compressibility of different regions in one data set, our framework can split the whole data set into multiple data blocks and calculate an entropy value for each block. Then, we can use this series of entropy values to identify in the given data set the regions that are harder to compressed.

4. Z-checker can estimate the smoothness of the original data sets by computing the first- and second-order partial derivatives along any dimension.
5. A multidimensional floating-point data set, $X = \{x_1, x_2, ..., x_N\}$, can be seen as a discrete signal. For a given signal, the power spectral density (or simply power spectrum) gives the power (energy per unit of time) of the frequencies present in the signal. The most common way of generating a power spectrum is using a DFT. The power spectrum transform is one kind of technique to transform the original data to the frequency domain. Many lossy compression algorithms, especially in image processing, often transform the original data to the frequency domain. For example, JPEG (Wallace, 1992) uses a discrete cosine transform, and JPEG2000 (Taubman and Marcellin, 2012) uses a DWT. Thus, our framework can evaluate the power spectrum of the original data using an efficient implementation of fast Fourier transform. Also, our framework can evaluate the wavelet transform on the original data set and can implement any other transform functions based on user demand.

6.  Some lossy compressors differentiate the dimensions of data, such as SZ (Di and Cappello, 2016; Tao et al., 2017) and ZFP (Lindstrom, 2014). Specifically, SZ uses a curve-fitting technique to predict the next data point only along the first dimension; it is not designed to be used along the other dimensions or use a dynamic selection mechanism for the dimension. Actually, the prediction accuracy depends on how much useful information is used in the prediction; hence, a good prediction should be performed along the dimension that contains the most useful information. The PCA technique is used to reduce the dimensionality of the data. Also, it may help identify which dimension contains the most information. Therefore, our framework adopts PCA for analyzing the original data set for guiding the design of lossy compression algorithm.

7.  To evaluate the correlation of neighbored data, our framework can analyze the property of autocorrelation for the original data sets. The autocorrelation coefficients AC of the original data set are calculated as follows

$$\text{AC}(\tau) = \frac{E[(x_i - \mu)(x_{i+\tau} - \mu)]}{\sigma^2} \quad (3)$$

where $\mu$ and $\sigma^2$, respectively, are the mean and covariance value of the original data set $X$, and $x_i$ and $x_{i+\tau}$ represent the data point $i$ and data point $i + \tau$, respectively; hence, $\tau$ means the spatial lag.

## 4.2. Assessment functions for compression checker of Z-checker

We now discuss the metrics and functions for assessing lossy compressors.

1.  For data point $i$, let $e_{\text{abs}_i} = x_i - \tilde{x}_i$, where $e_{\text{abs}_i}$ is the *absolute error*; let $e_{\text{rel}_i} = e_{\text{abs}_i}/R_X$, where $e_{\text{rel}_i}$ is the *value range–based relative error*. To evaluate the pointwise difference between the original and compressed data, our framework can compute the maximum absolute error $e_{\text{abs}}^{\max}$ and maximum value range–based relative error $e_{\text{rel}}^{\max}$ for all the data points, for example, $e_{\text{abs}}^{\max} = \max_{1 \leq i \leq N} e_{\text{abs}_i}$ and $e_{\text{rel}}^{\max} = \max_{1 \leq i \leq N} e_{\text{rel}_i}$. For error-controlled lossy compression, the compression errors will be guaranteed within the error bounds, which can be expressed by the formula $|e_{\text{abs}}^{\max}| < eb_{\text{abs}}$ or/and $|e_{\text{rel}}^{\max}| < eb_{\text{rel}}$, where $eb_{\text{abs}}$ is the absolute error bound and $eb_{\text{rel}}$ is the value range–based relative error bound.

2.  To evaluate the average error in the compression, our framework can compute the RMSE.

$$\text{RMSE} = \sqrt{\frac{1}{N} \sum_{i=1}^{N} (e_{\text{abs}_i})^2} \quad (4)$$

Because of the diversity of variables, our framework can further calculate the NRMSE.

$$\text{NRMSE} = \frac{\text{RMSE}}{R_X} \quad (5)$$

The PSNR is another commonly used average error metric for evaluating a lossy compression method, especially in visualization; hence, our framework also analyzes the PSNR. It is calculated as follows

$$\text{PSNR} = -20 \cdot \log_{10}(\text{NRMSE}) \quad (6)$$

The PSNR measures the size of the RMSE relative to the peak size of the signal. Logically, a lower value of RMSE/NRMSE means less error, but a higher value of PSNR represents less error.

3.  To evaluate the distribution of compression errors, our framework first loads the original data set based on the application name and its corresponding variable name kept in a hashtable, and then queries the decompressed data sets based on the compressor name and a specific compression error. After that, Z-checker computes the set of pointwise compression errors and generates the distribution of errors by calling the PDF computation function.

4.  To evaluate the size reduction as a result of the compression, our framework calculates the CR

$$\text{CR}(F) = \frac{\text{filesize}(F_{\text{orig}})}{\text{filesize}(F_{\text{comp}})} \quad (7)$$

or the BR (bits/value)

$$\text{BR}(F) = \frac{\text{filesize}_{\text{bit}}(F_{\text{comp}})}{N} \quad (8)$$

where filesize$_{\text{bit}}$ is the file size in bits and $N$ is the data size. The BR represents the amortized storage cost of each value. For a single/double floating-point data set, the BR is 32/64 bits per value before a compression, while the BR will be less than 32/64 bits per value after a compression. Also, CR and BR have a mathematical relationship as $\text{BR}(F) \times \text{CR}(F) = 32/64$; hence, a lower BR means a higher CR.

5.  Some compressors are designed for a fixed BR, such as ZFP, whereas some compressors are designed for a fixed maximum compression error, such as SZ and ISABELA. Our framework can plot the *rate–distortion* curves for different compressors to compare the distortion quality with the same rate. Here, rate means BR in bits/value, and our framework uses the PSNR to measure the distortion quality. PSNR is calculated by equation (6) in decibels. Generally speaking, in the rate–distortion curve, the higher the BR (i.e. more bits per value) in compressed data, the higher the quality (i.e. higher PSNR) of the reconstructed data after decompression. To plot the rate–distortion for a specific lossy

**Table 3.** Descriptions of preliminary test data sets evaluated by Z-checker.

| Name | Domain | Description |
|---|---|---|
| CESM | Simulation | CAM-SE cubed sphere atmosphere (ATM) simulation from CESM |
| Hurricane | Simulation | Hurricane Isabel simulation produced by WRF model |
| HACC | Simulation | Next-generation dark matter cosmology simulations based on HACC |
| Miranda | Simulation | Radiation hydrodynamics code designed for large-eddy simulation of multicomponent flows with turbulent mixing |
| EXAALT | Simulation | Exascale molecular dynamics simulation for spanning the accuracy, length, and time scales in materials science |
| APS | Instrument | Next-generation APS project for high-energy (hard) X-ray beams |
| EXAFEL | Instrument | Exascale modeling of advanced particle accelerators |

CESM: Community Earth System Model; WRF: Weather Research and Forecast; HACC: Hybrid/Hardware Accelerated Cosmology Code; APS: Advanced Photon Source.

compressor, Z-checker performs multiple compressions for various error bounds (such as $10^{-7}$, $10^{-6}$, and $10^{-5}$, which can be adjusted by the user in the configuration file), and then constructs the rate–distortion figure by checking the corresponding PSNRs in increasing order of the compression rates.

6. To evaluate the correlation between original and reconstructed data sets, our framework adopts the Pearson correlation coefficient $\rho$

$$\rho = \frac{\text{cov}(X, \tilde{X})}{\sigma_X \sigma_{\tilde{X}}} \qquad (9)$$

where $\text{cov}(X, \tilde{X})$ is the covariance. This coefficient is a measurement of the linear dependence between two variables, giving $\rho$ between $+1$ and $-1$, where $\rho = 1$ is the total positive linear correlation. The APAX profiler (Wegener, 2013) suggests that the correlation coefficient between original and reconstructed data should be 0.99999 ("five nines") or better.

7. To evaluate the speed of compression, our framework analyzes the throughput (bytes per second) based on the execution time of both compression and decompression process.

8. Since some applications require the compression errors to be uncorrelated (Wu and Huang, 2004), our framework analyzes the autocorrelation of the compression errors. The autocorrelation coefficients of the compression errors are calculated as

$$\text{ACE}(\tau) = \frac{E[(e_{\text{abs}i} - \mu)(e_{\text{abs}(i+\tau)} - \mu)]}{\sigma^2} \qquad (10)$$

where $\mu$ and $\sigma^2$ represent the mean and covariance value of the compression errors $e_{\text{abs}i}$, respectively. Similar to the autocorrelation of the original data sets, $\tau$ also means the spatial delay. The absolute uncorrelated compression errors require the autocorrelation coefficients to be 0 for all $\tau$.

9. Lossy compression typically is used for storing the original data, but some simulations need calculating derivatives, for example, first or second derivatives, divergence (i.e. sum of the first-order partial derivatives), or Laplacian (i.e. sum of the second-order partial derivatives). However, the computations of derivatives may enlarge any errors introduced by lossy compression, and the deviation is not readily apparent in the original data. For example, enlargement of the derived data may result in artifacts in visualization (ZFP and Derivatives, 2016). Higher-order derivatives are more sensitive to the compression errors. Thus, our framework compares several derivatives on the original and reconstructed data sets for lossy compression methods, such as first- or second-order partial derivatives, divergence, and Laplacian. For a three-dimensional (3-D) data set, we estimate divergence (div) and Laplacian (Lap) using central differencing as follows

$$\begin{aligned}
\text{div}(x,y,z) &= f_x + f_y + f_z \\
&= f(x-1,y,z) + f(x,y-1,z) + f(x,y,z-1) \\
&\quad -3f(x,y,z), \\
\text{Lap}(x,y,z) &= f_{xx} + f_{yy} + f_{zz} \\
&= f(x-1,y,z) + f(x+1,y,z) + f(x,y-1,z) \\
&\quad + f(x,y+1,z) + f(x,y,z-1) + f(x,y,z+1) \\
&\quad -3f(x,y,z),
\end{aligned}$$

where $f(x,y,z)$ represents the data value of location $(x,y,z)$ in the original field. For more straightforward comparison, our framework can also calculate RMSE, NRMSE, and PSNR based on the derived fields of the original data set and the reconstructed data set.

10. Because of information loss, lossy compression may affect data transforms. Hence, our framework can compare the results of data transforms on the original and reconstructed data sets. For example, our framework is able to evaluate the effects that lossy compressors bring to DFT. It can generate two sets of DFT results on the original and reconstructed data sets at the same time and compare the amplitudes of these two DFT results from low frequency to high frequency. Additionally, the framework is implemented with assessing lossy compressors under DWT, and users can also implement any other data transforms into the tool.

## 5. Diverse usage of Z-checker

In this section, we first present an assessment of the lossy compression qualities of several lossy compressors (JPEG2000, ISABELA, FPZIP, ZFP, and SZ) using Z-checker analysis. Note that we refer to the improved version of the SZ lossy compressor, that is, SZ-1.4

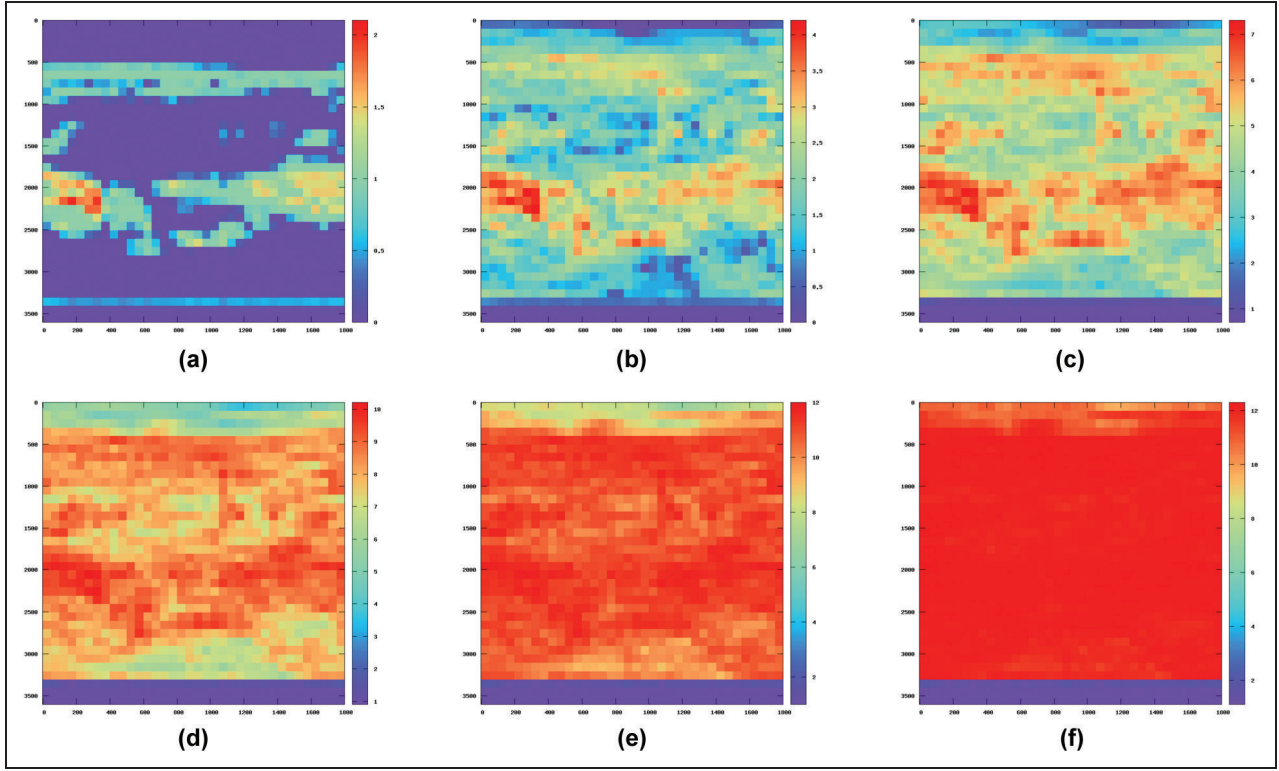

**Figure 3.** Z-checker visualization of the entropy (block) with different accuracies on CESM data sets. (a) $eb_{abs} = 10^{-1}$; (b) $eb_{abs} = 10^{-2}$; (c) $eb_{abs} = 10^{-3}$; (d) $eb_{abs} = 10^{-4}$; (e) $eb_{abs} = 10^{-5}$; (f) $eb_{abs} = 10^{-6}$. CESM: Community Earth System Model.

**Table 4.** Evaluation of maximum compression error with different lossy compressors and user-set $eb_{rel}$ on CESM and Hurricane data sets visualized by Z-checker.

|  | CESM | | Hurricane | |
| --- | --- | --- | --- | --- |
| User-set $eb_{rel}$ | SZ | ZFP | SZ | ZFP |
| $10^{-2}$ | $1.0 \times 10^{-2}$ | $3.3 \times 10^{-3}$ | $1.0 \times 10^{-2}$ | $2.2 \times 10^{-3}$ |
| $10^{-3}$ | $1.0 \times 10^{-3}$ | $4.3 \times 10^{-4}$ | $1.0 \times 10^{-3}$ | $1.4 \times 10^{-3}$ |
| $10^{-4}$ | $1.0 \times 10^{-4}$ | $2.6 \times 10^{-5}$ | $1.0 \times 10^{-4}$ | $1.8 \times 10^{-5}$ |
| $10^{-5}$ | $1.0 \times 10^{-5}$ | $3.4 \times 10^{-6}$ | $1.0 \times 10^{-5}$ | $2.1 \times 10^{-6}$ |
| $10^{-6}$ | $1.0 \times 10^{-6}$ | $4.1 \times 10^{-7}$ | $1.0 \times 10^{-6}$ | $2.8 \times 10^{-7}$ |

CESM: Community Earth System Model.

(Tao et al., 2017), as SZ in the following sections for simplification. For this assessment, we consider various scientific simulation and instrument data sets including climate simulation (CESM), cosmology simulation (HACC), radiation hydrodynamics simulation (Miranda), molecular dynamics simulation (EXAALT), and X-ray laser instruments (APS and EXAFEL). Table 3 shows the descriptions of the preliminary test data sets in detail. Most of them are from the applications being developed in the ECP.

### 5.1. Using Z-checker to assess lossy compression quality

In this subsection, we first show the experimental results assessed by our framework Z-checker from different aspects, including data analyzer and compression checker. Then, we explain each analysis and assessment result and give some suggestions to the selection or design of the lossy compressors.

*5.1.1. Entropy.* The entropy is an indicator of the compressibility. We can use it to identify in a given data set the regions that are harder to compress. Figure 3 shows the visualization results of block entropy values with different accuracies ($10^{-1}$ to $10^{-6}$). The higher the entropy value, the harder it is to compress the data set; hence, we are able to use this property to guide which areas are hard to compress. For example, in Figure 3, we can infer that the red area is harder to compress than the green and purple area. If the accuracy (or absolute error bound) is set to $10^{-3}$, an efficient lossy compressor should focus mainly on the data bands in red color located in the center region. Note that in Figure 3, the block size for entropy computation is $100 \times 100$ and the original data set is $1800 \times 3600$.

*5.1.2. Maximum compression error.* Table 4 shows the maximum compression error with different lossy compressors evaluated by Z-checker. Table 4 shows that with the same user-set error bound, the maximum compression error of ZFP is lower than that of SZ. ZFP overpreserves the original data with respect to the user-set error bound (or accuracy). Thus, our framework may guide developers of ZFP compressor to relax the compression error, thereby improving the CR.

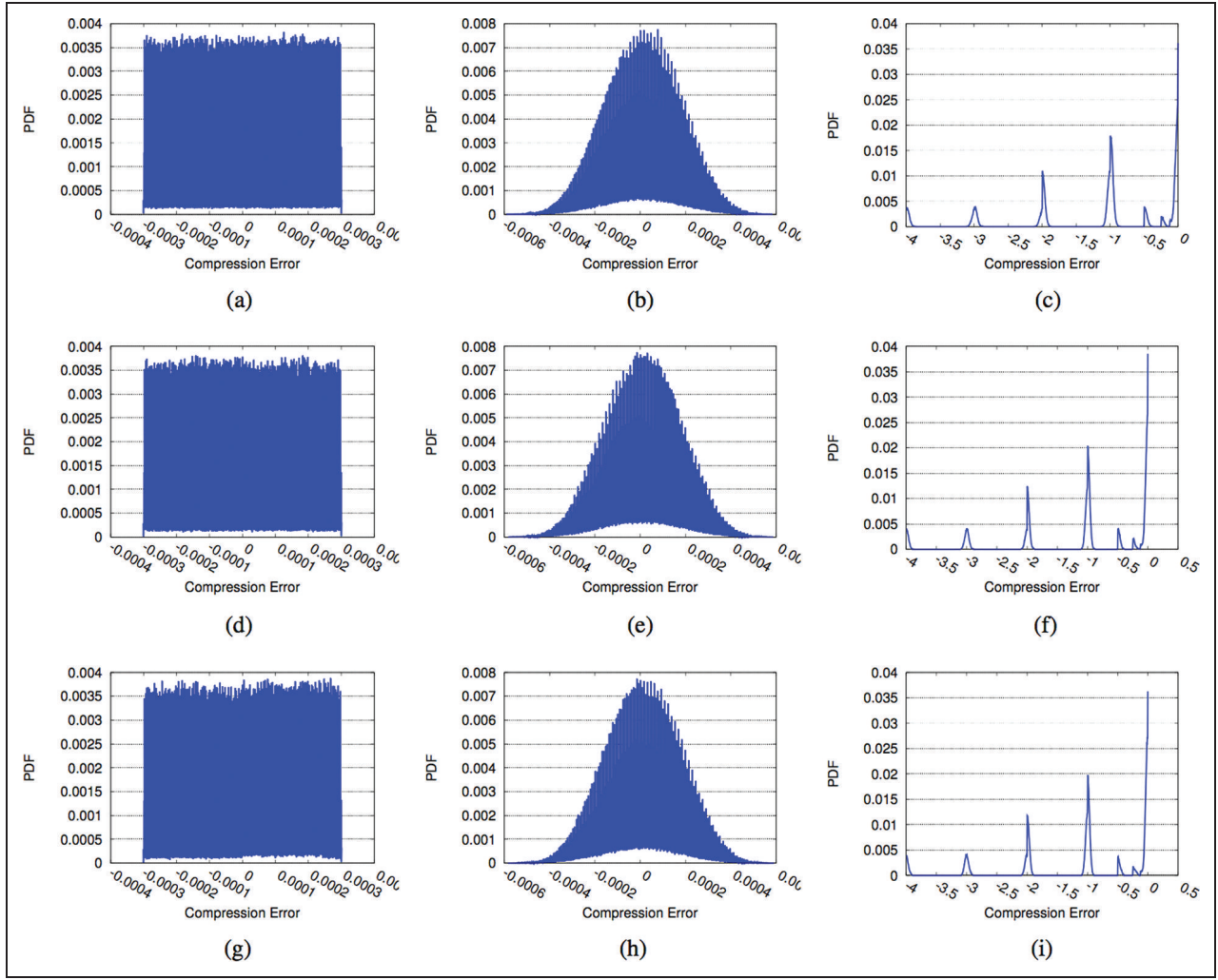

**Figure 4.** Evaluation of compression error distribution (PDF) with different lossy compressors on HACC data sets (variables *vx*, *vy*, and *vz*) visualized by Z-checker. (a) SZ(*xx*); (b) ZFP(*xx*); (c) FPZIP(*xx*); (d) SZ(*yy*); (e) ZFP(*yy*); (f) FPZIP(*yy*); (g) SZ(*zz*); (h) ZFP(*zz*); and (i) FPZIP(*yy*). PDF: probability density function; HACC: Hardware/Hybrid Accelerated Cosmology Code.

*5.1.3. Distribution of compression errors.* As an example, Figure 4 presents the evaluation results of the distributions of compression errors, produced by Z-checker, with different lossy compressors on the HACC data sets. In particular, we can see that the distribution of SZs compression error is nearly *uniform*, while ZFP is nearly *normal*. In this sense, the users are able to select the appropriate compressors in terms of their expected error distribution (either uniform or normal distribution in this case), based on the analysis results produced by Z-checker.

*5.1.4. Compression ratio.* Figure 5 shows the CRs evaluated by Z-checker with different lossy and lossless compressors, including SZ, ZFP, ISABELA, FPZIP, and GZIP, on the CESM, APS, and Hurricane data sets. Figure 6 presents the CRs evaluated by Z-checker with different lossy compressors on HACC data sets. Note that in Figure 6, SZ evaluated two compression modes for SZ, including SZ_DEFAULT_COMPRESSION (denoted by SZ(d), where d indicates "default"), and SZ_BEST_SPEED (denoted by SZ(f), where f here refers to "fast"). FPZIP(24/12) means that the number of bits is set to 24 for particles' coordinate variables *xx*, *yy*, and *zz* and to 12 for their velocity variables *vx*, *vy*, and *vz* because these two configurations can get the error bound to be $10^{-3}$. Similarly, FPZIP(28/18) means that the number of bits is 28 for particles' coordinate variables *xx*, *yy*, and *zz* and the number of bits is set to 18 for their velocity variables *vx*, *vy*, and *vz* in order to get the error bound to be $10^{-5}$.

*5.1.5. Rate–distortion.* Figure 7 illustrates the results of rate–distortion produced by Z-checker for different lossy compressors on the HACC data sets. Note that SZs compression result is a little inferior to that of FPZIP only when the BR is around 16 (bits/value). The type of HACC data is single floating-point; hence, a BR of 16 bits per value means a CR of 2, which is nearly the CR of lossless compression (Ratanaworabhan et al., 2006) and far more precise than the user's requirements in general.

Figure 8 shows the results of rate–distortion produced by Z-checker for the SZ and ZFP compressors on the

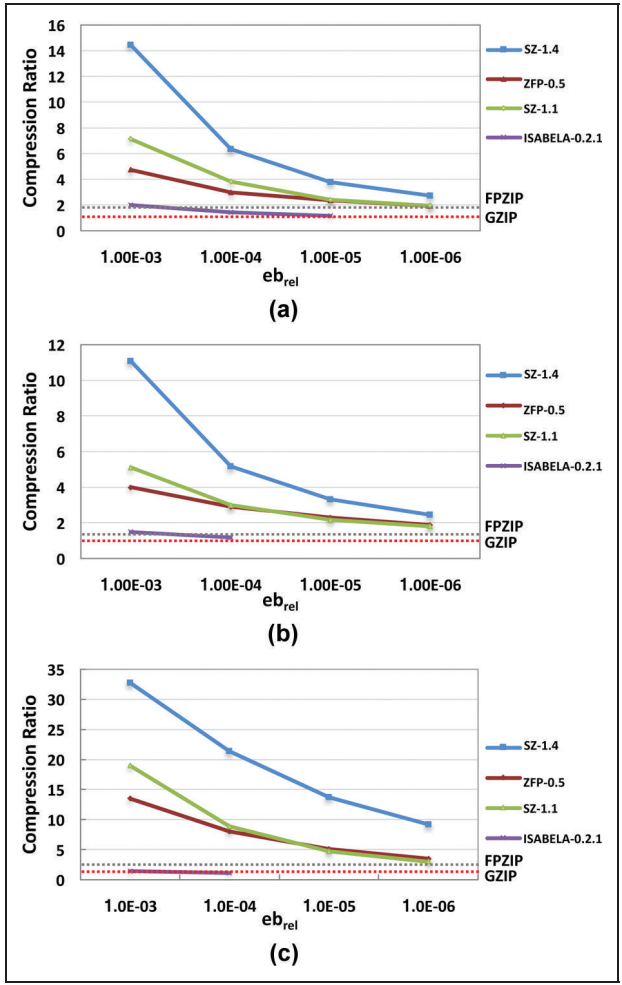

**Figure 5.** Evaluation of compression ratio with different lossy compressors on (a) CESM, (b) APS, and (c) Hurricane data sets visualized by Z-checker. CESM: Community Earth System Model; APS: Advanced Photon Source.

Hurricane and Miranda data sets. It illustrates that ZFP outperforms SZ on the Miranda data sets, whereas SZ outperforms ZFP on the Hurricane data sets. Both data sets are 3D. The reason ZFP has better performance on the Miranda data sets may be that (1) the coefficients of the transform matrix in ZFP were optimized for the Miranda data sets and (2) the accuracy of data prediction of SZ on the Miranda data sets is not as good as on the other data sets. Based on Figure 8, we see that SZ and ZFP exhibit better compression results than the other one does in different cases with a 3D array. Specifically, for Miranda simulations, Z-checker suggests using ZFP instead of SZ; developers of the SZ compressor can further optimize the SZ compressor in terms of rate–distortion on the Miranda data sets and more generic high-dimensional data sets. More results evaluated by Z-checker with respect to the rate–distortion are shown in Figure 9.

*5.1.6. Speed (or processing rate).* Figures 10 and 11 present the speeds of compression and decompression for different lossy compressors on the HACC data sets, respectively. Users of lossy compressors can use the speed of compression/decompression to estimate whether they can get a performance gain (save time) from lossy compression. Specifically, suppose that users want to transmit $D$ bytes data on networks or store $D$ bytes data in disks/PFS. Let us assume that the transmission bandwidth of networks or storage bandwidth of disks/PFS is BW and the compression/decompression speed is $R_{comp}/R_{decomp}$ (bytes per second). Also, let us assume that the CR using a certain lossy compressor is CR. The time consumed in transmitting/storing the original data is $D$/BW in seconds, while the time consumed in transmitting/storing the compressed data is $D$/CR BW in seconds. Also, we need to consider the overhead for compressing/decompressing the data; hence, the total time is $D/(\text{CR} \cdot \text{BW}) + D/R_{\text{comp}} + D/R_{\text{decomp}}$. Thus, users can benefit from lossy compression if

$$D/(\text{CR} \cdot \text{BW}) + D/R_{\text{comp}} + D/R_{\text{decomp}} < D/\text{BW}$$

which is equivalent to

$$\text{BW}/R_{\text{overall}} < (\text{CR} - 1)/\text{CR} \quad (11)$$

where $R_{\text{overall}}$ is the overall speed of once compression process and once decompression process, that is, $R_{\text{comp}} \cdot R_{\text{decomp}}/(R_{\text{comp}} + R_{\text{decomp}})$.

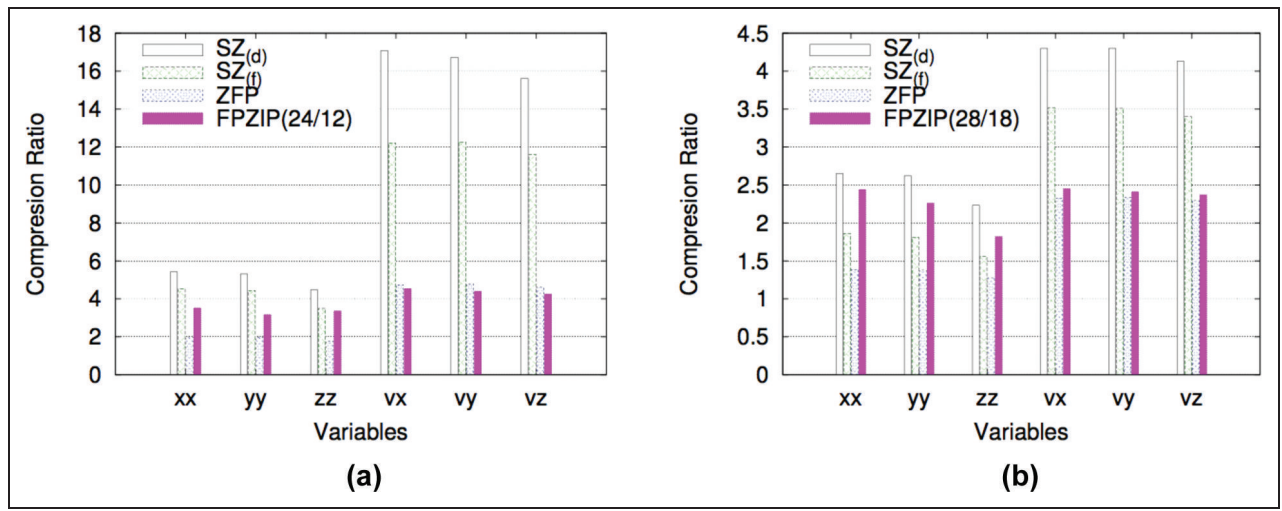

**Figure 6.** Evaluation of compression ratio with different lossy compressors on HACC data sets visualized by Z-checker. (a) $eb_{abs} = 10^{-3}$ and (b) $eb_{abs} = 10^{-5}$. HACC: Hardware/Hybrid Accelerated Cosmology Code.

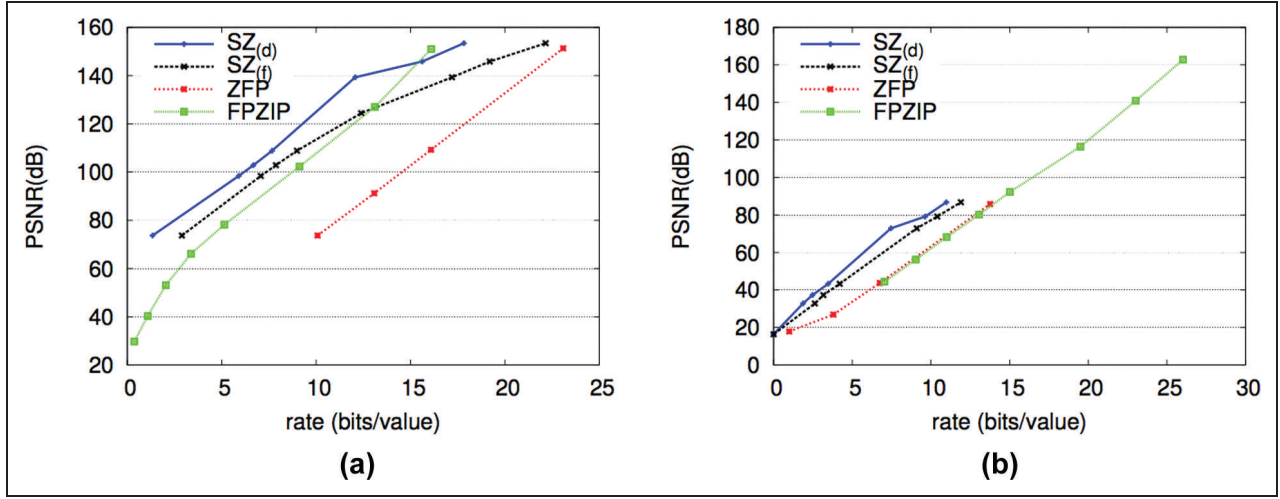

**Figure 7.** Evaluation of Z-checker for rate–distortion with different lossy compressors on HACC data sets displayed by Z-checker. (a) *xx* and (b) *vx*. HACC: Hardware/Hybrid Accelerated Cosmology Code.

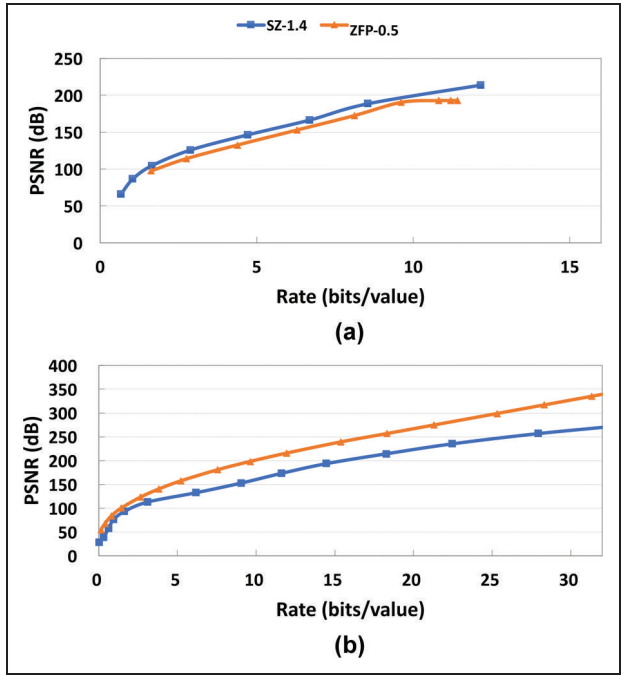

**Figure 8.** Evaluation of rate–distortion with different lossy compressors on 3D (a) Hurricane and (b) Miranda data sets visualized by Z-checker.

From equation (11), users can estimate how much time they can save from lossy compression. Note that Figures 10 and 11 show the compression/decompression speeds of different lossy compressors with only one process, while the compression/decompression rate in equation (11) is with multiple processes in a large-scale parallel environment.

*5.1.7. Pearson correlation.* Table 5 shows the Pearson correlation coefficients evaluated by Z-checker for different lossy compressors with different maximum compression errors on the CESM and Hurricane data sets. The APAX profiler (Wegener, 2013) suggests that the correlation coefficient between original and reconstructed data should be 0.99999 ("five nines") or better. Thus, to satisfy this suggestion, we should set the value range–based error bound to be around $10^{-4}$ or lower for the SZ and ZFP compressors.

*5.1.8. Autocorrelation of compression errors.* Figure 12 shows the evaluation results from Z-checker of the autocorrelation of the compression errors on two typical variables in the CESM data sets, that is, FREQSH and SNOWHLND. The CRs of FREQSH and SNOWHLND are 6.5 and 48 using SZ-1.4 with $eb_{\text{rel}} = 10^{-4}$. Thus, to some extent, FREQSH can represent relatively low-compression factor data sets, while SNOWHLND can represent relatively high-compression factor data sets. The figure illustrates that on the FREQSH, the maximum autocorrelation coefficient of SZ-1.4 is $4 \times 10^{-3}$, which is much lower than ZFPs 0.25. However, on the SNOWHLND, the maximum autocorrelation coefficient of SZ-1.4 is about 0.5, which is higher than ZFPs 0.23. It illustrates that the effect of compression error autocorrelation being application specific, lossy compressor users might need to understand this effect before using one of the other compressor.

*5.1.9. Comparison of data transform on original and reconstructed data.* Figure 13 shows the comparison of DFT on the calibrated variable in the EXAFEL data sets evaluated by Z-checker. Z-checker can generate two sets of DFT

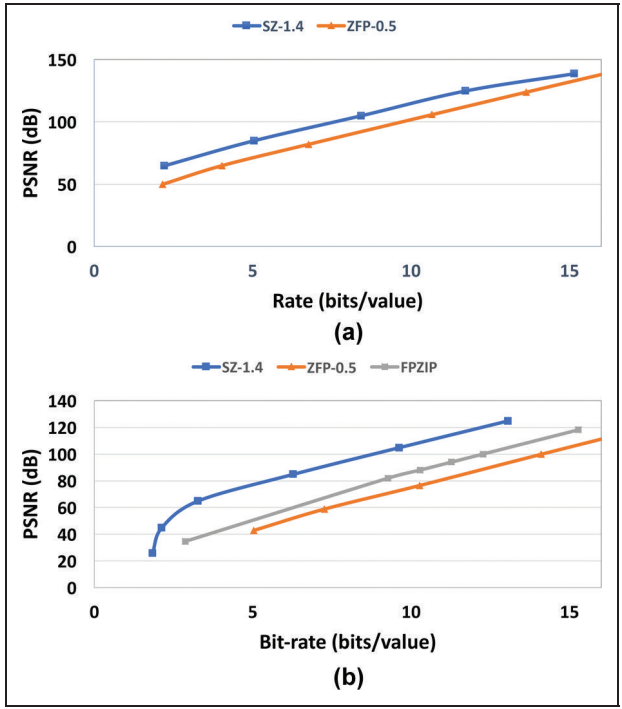

**Figure 9.** Evaluation of rate–distortion with different lossy compressors on 2D (a) CESM and (b) EXAALT data sets visualized by Z-checker. CESM: Community Earth System Model.

results on the original and reconstructed EXAFEL data sets at the same time and compare the amplitudes of these two DFT results from low frequency to high frequency. The figure shows the difference of these two DFTs amplitudes using SZ and JPEG2000 on one panel in the calibrated data sets. For SZ, Z-checker tests two error bounds, which are the value range–based relative error bound $10^{-2}$ and $10^{-3}$. For JPEG2000, Z-checker uses the CRs as same as SZs. Note that the difference is normalized to each amplitude of the DFT performed on the original data sets; hence, this difference can be considered as pointwise relative error of

DFTs amplitudes. The figure illustrates that the amplitude differences of SZ with the value range–based relative error bound $10^{-3}$ are smaller than the amplitude differences of JPEG2000 with the corresponding CRs. Specifically, the normalized difference of amplitudes using SZ can be within 1% on the calibrated data set, whereas, the normalized difference of amplitudes using JPEG2000 can only be within 500% with most of normalized differences are within 100%.

## 6. Related work

Baker et al. (2014) investigated the use of data compression techniques on climate simulation data from CESM (2017). They developed an approach for verifying the climate data and used it to evaluate several compression algorithms, including FPZIP, ISABELA, APAX, and GRIB2 (with JPEG2000 compression). The verification process included (1) quantifying the difference between the original and reconstructed data sets via measures of pointwise error, average error (RMSE and NRMSE), and Pearson correlation and (2) evaluating the reconstructed data in the context of an ensemble of CESM runs with slight perturbations by a CESM port verification tool (CESM-PVT). Specifically, CESM-PVT can determine which decompressed/reconstructed variables are good to use through performing CESM-PVT test on an ensemble consisting of multiple climate simulations. Also, CESM-PVT can evaluate whether lossy compression has added any bias to the climate data. Moreover, CESM-PVT can evaluate whether the maximum pointwise compression error between the original and reconstructed data is reasonable. They determined that the diversity of the climate data requires individual treatment of variables and that the reconstructed data can fall within the natural variability of the system. Laney et al. (2014) examined the effects of lossy compression in physics simulations by evaluating two lossy compressors (FPZIP and APAX) in three physics simulation codes. They

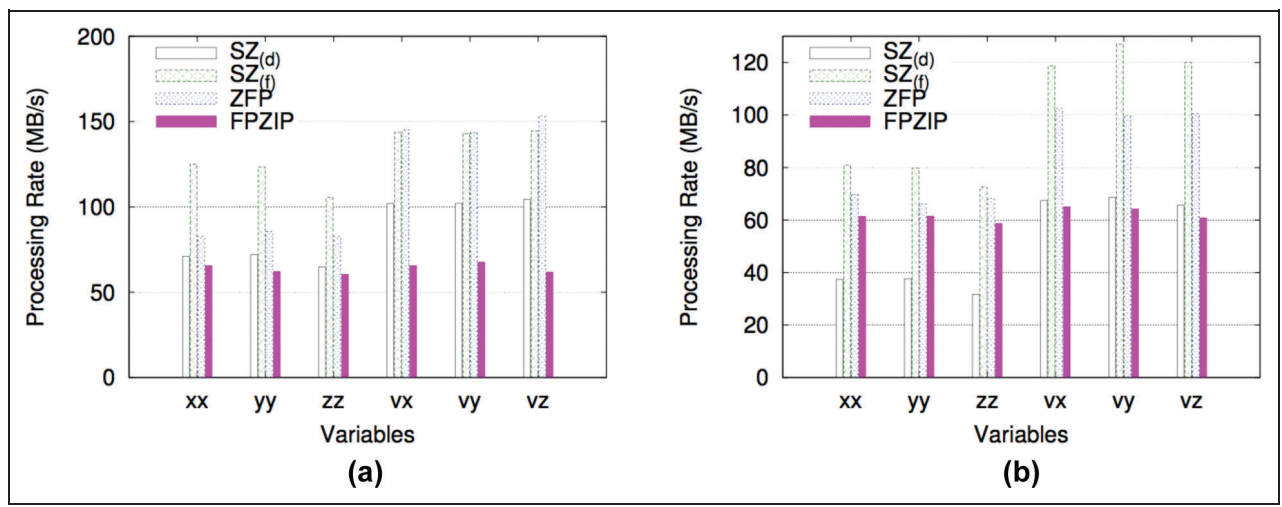

**Figure 10.** Evaluation of compression speed with different lossy compressors on HACC data sets visualized by Z-checker. (a) $eb_{abs} = 10^{-3}$ and (b) $eb_{abs} = 10^{-5}$. HACC: Hardware/Hybrid Accelerated Cosmology Code.

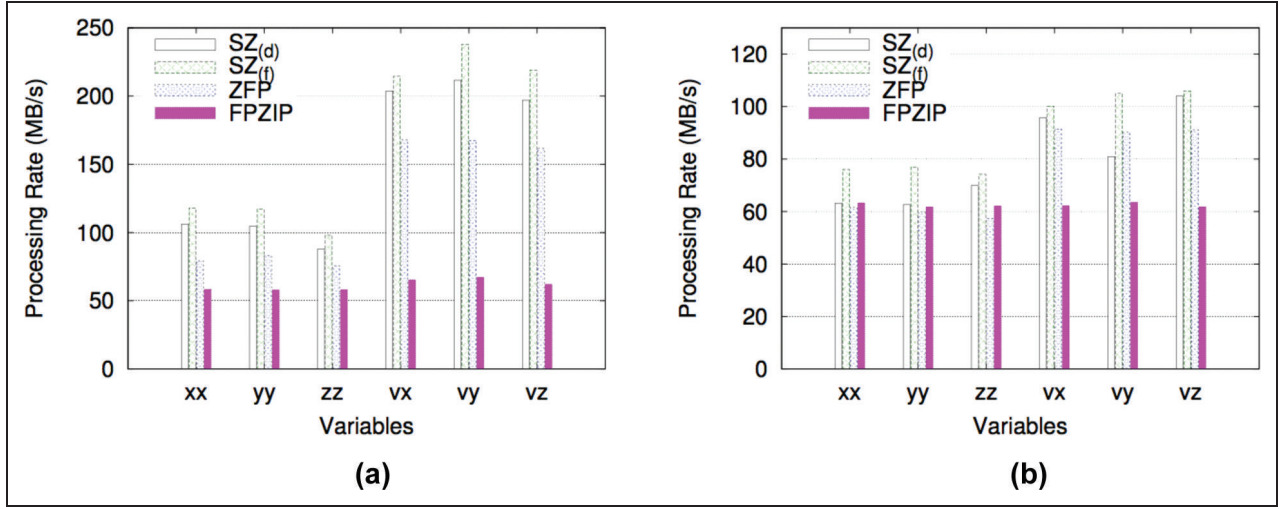

**Figure 11.** Evaluation of decompression speed with different lossy compressors on HACC data sets displayed by Z-checker. (a) $eb_{abs}$ = $10^{-3}$ and (b) $eb_{abs}$ = $10^{-5}$. HACC: Hardware/Hybrid Accelerated Cosmology Code.

**Table 5.** Evaluation of Pearson correlation with different lossy compressors on CESM and Hurricane data sets visualized by Z-checker.

| | CESM | | | Hurricane | |
|---|---|---|---|---|---|
| Maximum $eb_{rel}$ | SZ | ZFP | Maximum $eb_{rel}$ | SZ | ZFP |
| $3.3 \times 10^{-3}$ | 0.9998 | 0.9996 | $2.2 \times 10^{-3}$ | 0.998 | 0.99995 |
| $4.3 \times 10^{-4}$ | $\geq 1 - 10^{-6}$ | $\geq 1 - 10^{-7}$ | $1.4 \times 10^{-4}$ | $\geq 1 - 10^{-5}$ | $\geq 1 - 10^{-6}$ |
| $2.6 \times 10^{-5}$ | $\geq 1 - 10^{-9}$ | $\geq 1 - 10^{-9}$ | $1.8 \times 10^{-5}$ | $\geq 1 - 10^{-6}$ | $\geq 1 - 10^{-8}$ |
| $3.4 \times 10^{-6}$ | $\geq 1 - 10^{-11}$ | $\geq 1 - 10^{-11}$ | $2.1 \times 10^{-6}$ | $\geq 1 - 10^{-8}$ | $\geq 1 - 10^{-9}$ |
| $4.1 \times 10^{-7}$ | $\geq 1 - 10^{-13}$ | $\geq 1 - 10^{-13}$ | $2.8 \times 10^{-7}$ | $\geq 1 - 10^{-10}$ | $\geq 1 - 10^{-11}$ |

CESM: Community Earth System Model.

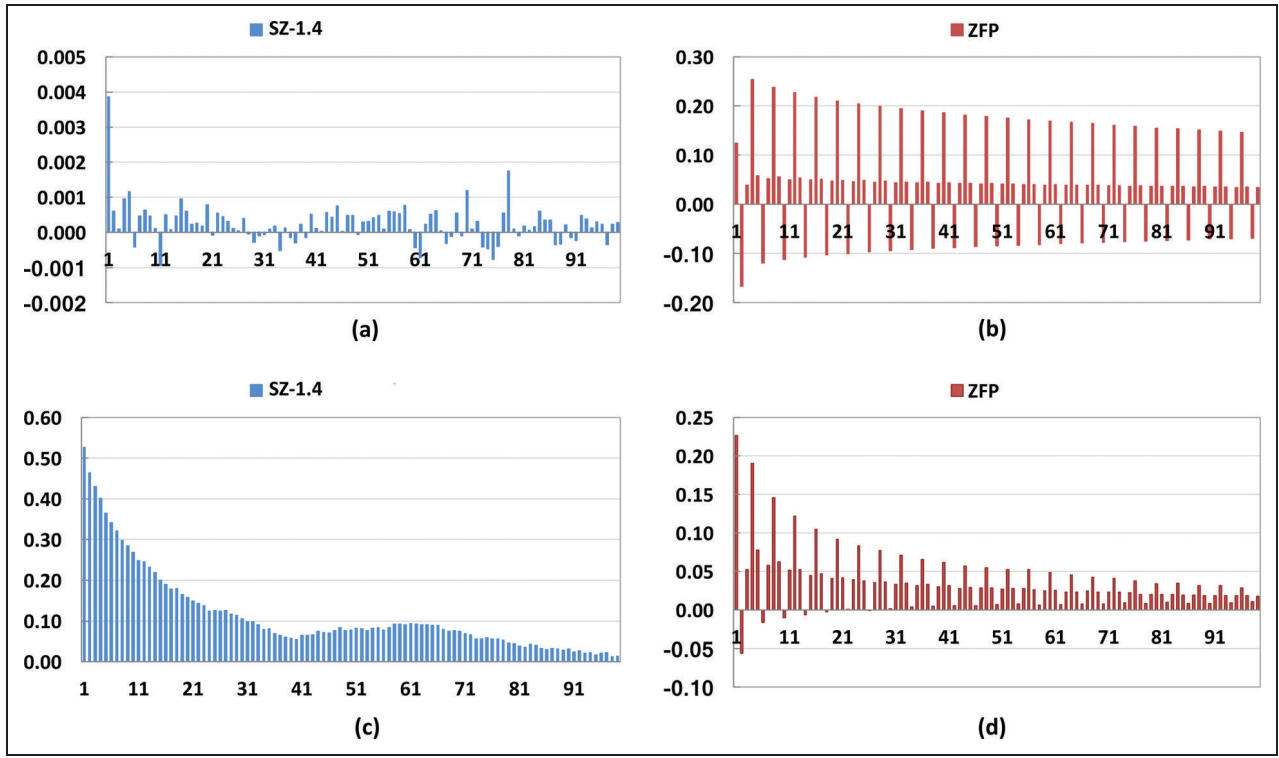

**Figure 12.** Evaluation of autocorrelation of compression errors with different delays (first 100) using (a) SZ and (b) ZFP on CESM and Hurricane data sets visualized by Z-checker. (a) FREQSH. (b) FREQSH. (c) SNOWHLND. (d) SNOWHLND. CESM: Community Earth System Model.

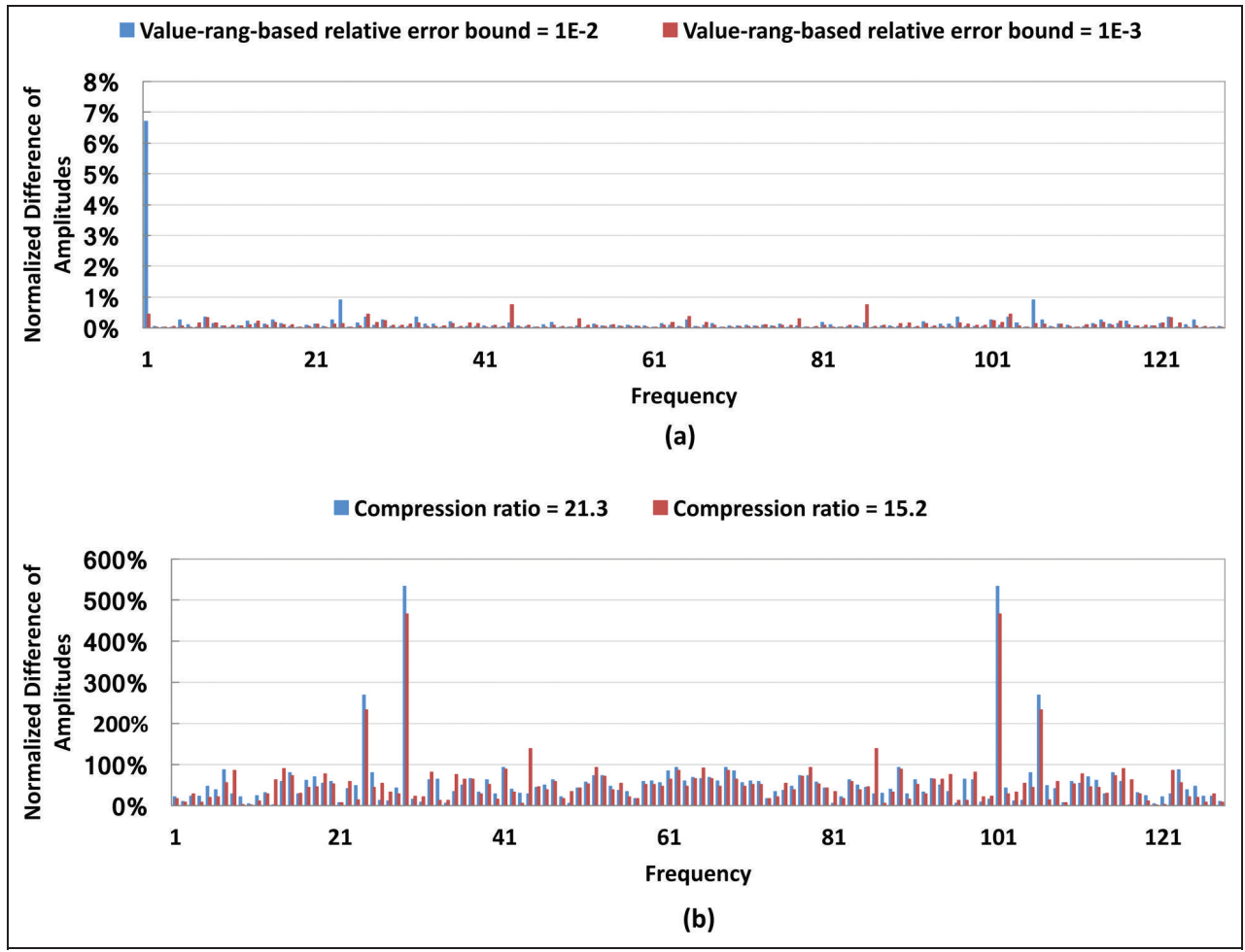

**Figure 13.** Evaluation of normalized difference of DFTs amplitudes on original and reconstructed calibrated variable in EXAFEL data sets with (a) SZ and (b) JPEG2000 visualized by Z-checker. DFT: discrete Fourier transform.

used physics-based metrics for each simulation to assess the impact of lossy compression. They noted that the characteristics of the compression error must be carefully considered in the context of the underlying physics being modeled. Lakshminarasimhan et al. (2013) proposed the ISABELA lossy compressor. It performs data compression by B-spline interpolation after sorting the data series. They evaluated the performance of ISABELA with several metrics: CR, maximum compression error (pointwise relative error), average compression error (NRMSE), and compression time. They also evaluated the compressed data by quantitative analysis and visual analysis. In quantitative analysis, they evaluated the Pearson correlation of different data regions and the difference of derived data. In visual analysis, they utilized the visualization tool to present the original data and ISABELA-compressed data. Lindstrom (2014) proposed a lossy compression algorithm for floating-point arrays in fixed rate (i.e. ZFP) and evaluated ZFP of rate–distortion, rate speed, density spectrum, and derivatives (Morse segmentation of gradients) in several applications, including quantitative and visual analysis, visualization, and fluid dynamics simulation. Sasaki et al. (2015) proposed a lossy compression method (i.e. SSEM) based on wavelet transform and VQ. They applied their compression method to a checkpoint/restart technique and evaluate the impact on results of a production climate application (NICAM). Di and Cappello (2016) proposed an error-bounded lossy compressor (i.e. SZ) based on curve-fitting and binary representation analysis. They evaluated SZ with the metrics including the maximum compression error (bounded or not bounded), CR, and compression/decompression speed. Tao et al. (2017) further improve the SZ lossy compressor (i.e. SZ-1.4) based on the multidimensional prediction model and uniform scalar quantization method. They evaluated SZ-1.4 with multiple metrics including compression error, CR, rate–distortion, Pearson correlation, compression/decompression speed, and autocorrelation of compression error.

## 7. Conclusion

In this article, we present a novel framework Z-checker for data analytics and lossy compression assessment. We carefully implement and will release it as an open-source library/software under a BSD license. The key functions of this framework are (1) exploring the properties of original data sets for data analytics or improvement of lossy compression algorithms, (2) assessing lossy compressors for specific scientific data sets, and (3) generating visual results by both static data visualization scripts and an interactive visualization system on demand. We evaluate our Z-checker based on various scientific simulation data sets across multiple state-of-art compressors, such as SZ, ZFP, FPZIP, ISABELA, JPEG2000, and GZIP. We also present how we improved SZ lossy compressor significantly for hard-to-compress data using the analysis results generated by Z-checker.

In the future work, we plan to (1) integrate more evaluation metrics and functions into the data property analyzer and compression checker of our framework, (2) finish the implementation of the online processing mode for Z-checker, (3) extend our frame to perform analysis dynamically and in situ to control the compression quality during the execution, and (4) implement specific assessment functions to handle the verification process for reconstructed data.

## Authors' note

The submitted manuscript has been created by UChicago Argonne, LLC, Operator of Argonne National Laboratory ("Argonne"). Argonne, a U.S. Department of Energy Office of Science laboratory, is operated under Contract No. DE-AC02-06CH11357. The U.S. Government retains for itself, and others acting on its behalf, a paid-up nonexclusive, irrevocable worldwide license in said article to reproduce, prepare derivative works, distribute copies to the public, and perform publicly and display publicly, by or on behalf of the Government.

## Funding

The author(s) disclosed receipt of the following financial support for the research, authorship, and/or publication of this article: This research was supported by the Exascale Computing Project (ECP), Project Number: 17-SC-20-SC, a collaborative effort of two DOE organizations – the Office of Science and the National Nuclear Security Administration, responsible for the planning and preparation of a capable exascale ecosystem, including software, applications, hardware, advanced system engineering and early testbed platforms, to support the nation's exascale computing imperative. The material also was supported in part by the U.S. Department of Energy, Office of Science, under contract DE-AC02-06CH11357. This material is also based upon work supported by the National Science Foundation under Grant No. 1619253.

## Notes

1. All three data loading ways described here are in the offline processing mode. Online processing will be implemented in our future work.
2. https://nodejs.org/.
3. https://github.com/highcharts/highcharts.
4. https://d3js.org/.
5. Note that unlike the pointwise relative error that is compared with each data value, value range–based relative error is compared with value range.

## References

Austin E (2016) Advanced photon source. *Synchrotron Radiation News* 29(2): 29–30.

Austin W, Ballard G and Kolda TG (2015) Parallel tensor compression for large-scale scientific data. In: *Parallel and Distributed Processing Symposium, 2016 IEEE International*, pp. 912–922. IEEE, 2016.

Baker AH, Hammerling DM, Mickelson SA, et al. (2016) Evaluating lossy data compression on climate simulation data within a large ensemble. *Geoscientific Model Development* 9(12): 4381.

Baker AH, Xu H, Dennis JM, et al. (2014) A methodology for evaluating the impact of data compression on climate simulation data. In: *Proceedings of the 23 rd international symposium on high-performance parallel and distributed computing*, pp. 203–214. ACM.

Bernholdt D, Bharathi S, Brown D, et al. (2005) The earth system grid: supporting the next generation of climate modeling research. *Proceedings of the IEEE* 93(3): 485–495.

Chen Z, Son SW, Hendrix W, et al. (2014) Numarck: machine learning algorithm for resiliency and checkpointing. In: *Proceedings of the International Conference for High Performance Computing, Networking, Storage and Analysis*, pp. 733–744. IEEE.

Community Earth Simulation Model (CESM) (2017) Available at: http:// www.cesm.ucar.edu/ (accessed 28 October 2017).

Deutsch LP (1996) Gzip file format specification version 4.3.

Di S and Cappello F (2016) Fast error-bounded lossy HPC data compression with SZ. In: *Parallel and Distributed Processing Symposium, 2016 IEEE International*, pp. 730–739. IEEE.

Gleckler PJ, Durack PJ, Stouffer RJ, et al. (2016) Industrial-era global ocean heat uptake doubles in recent decades. *Nature Climate Change* 6(4): 394–398.

Habib S, Morozov V, Frontiere N, et al. (2016) HACC: extreme scaling and performance across diverse architectures. *Communications of the ACM* 60(1): 97–104.

Huffman DA, et al. (1952) A method for the construction of minimum-redundancy codes. *Proceedings of the IRE* 40(9): 1098–1101.

Janert PK (2009) *Gnuplot in Action*. Greenwich: Manning Publications Co. ISBN: 1–933988.

Lakshminarasimhan S, Shah N, Ethier S, et al. (2013) ISABELA for effective in situ compression of scientific data. *Concurrency and Computation: Practice and Experience* 25(4): 524–540.

Laney D, Langer S, Weber C, et al. (2014) Assessing the effects of data compression in simulations using physically motivated metrics. *Scientific Programming* 22(2): 141–155.

Lindstrom P (2014) Fixed-rate compressed floating-point arrays. *TVCG* 20(12): 2674–2683.

Lindstrom P and Isenburg M (2006) Fast and efficient compression of floating-point data. *TVCG* 12(5): 1245–1250.

Ratanaworabhan P, Ke J and Burtscher M (2006) Fast lossless compression of scientific floating-point data. In: *Data Compression Conference 2006. DCC 2006 Proceedings*, pp. 133–142. IEEE.

Sasaki N, Sato K, Endo T, et al. (2015) Exploration of lossy compression for application-level checkpoint/restart. In: *Parallel and Distributed Processing Symposium (IPDPS), 2015 IEEE International*, pp. 914–922. IEEE.

Tao D, Di S, Chen Z, et al. (2017) Significantly improving lossy compression for scientific data sets based on multi-dimensional prediction and error-controlled quantization. In: *2017 IEEE international parallel and distributed processing symposium, IPDPS 2017*, Orlando, Florida, USA, 29 May–2 June, 2017.

Taubman D and Marcellin M (2012) *JPEG2000 Image Compression Fundamentals, Standards and Practice: Image Compression Fundamentals, Standards and Practice*, Vol. 642. Springer Science & Business Media.

Wallace GK (1992) The JPEG still picture compression standard. *IEEE Transactions on Consumer Electronics* 38(1): 28–34.

Wegener AW and Samplify Systems LLC (2006) *Adaptive Compression and Decompression of Bandlimited Signals*. U.S. Patent 7,009,533.

Wu Z and Huang NE (2004) A study of the characteristics of white noise using the empirical mode decomposition method. In: *Proceedings of the Royal Society of London A: Mathematical, Physical and Engineering Sciences* Vol. 460, No. 2046, pp. 1597–1611. The Royal Society.

ZFP and Derivatives (2016) Available at: http://computation.llnl.gov/projects/floating-point-compression/ zfp-and-derivatives (accessed 28 October 2017).

Ziv J and Lempel A (1977) A universal algorithm for sequential data compression. *IEEE Transactions on Information Theory* 23(3): 337–343.

Zuras D, Cowlishaw M, Aiken A, et al. (2008) IEEE standard for floating-point arithmetic. *IEEE Std 754-2008* 1–70.

## Author biographies

*Dingwen Tao* is a fifth-year doctoral candidate in computer science at the University of California, Riverside, under the advisement of Dr. Zizhong Chen. He received his bachelor degree in Information and Computing Science from the University of Science and Technology of China. He is currently working at Argonne National Laboratory in the Extreme Scale Resilience Group lead by Dr Franck Cappello. Prior to this, he worked in the High Performance Computing Group at Pacific Northwest National Laboratory in summer 2015. His research interests include high-performance computing, parallel and distributed computing, big data analytics, resilience and fault tolerance, data compression algorithms and softwares, numerical algorithms and softwares, and high-performance computing on heterogeneous systems. He has published 10+ peer-reviewed papers in top HPC and parallel and distributed conferences during his Ph.D. program, such as HPDC, IPDPS, PPoPP, SC. E-mail: dtao001@cs.ucr.edu.

*Sheng Di* received his master's degree from the Huazhong University of Science and Technology in 2007 and PhD degree from the University of Hong Kong in 2011. He is currently working at Argonne National Laboratory. His research interest involves resilience on high-performance computing (such as silent data corruption, optimization of checkpoint model, characterization and analysis of supercomputing log, and in situ data compression) and broad research topics on cloud computing (including optimization of resource allocation, cloud network topology, and prediction of cloud workload/host-load). He is the author of 17 papers published in international journals and 37 papers published at international conferences. He served as a programming committee member 10+ times for different conferences and served as an external conference/journal reviewer over 50 times. E-mail: sdi1@anl.gov.

*Hanqi Guo* is a postdoctoral appointee in the Mathematics and Computer Science Division, Argonne National Laboratory. He received his PhD degree in computer science from Peking University in 2014 and the BS degree in mathematics and applied mathematics from the Beijing University of Posts and Telecommunications in 2009. His research interests are mainly in large-scale scientific data visualization. E-mail: hguo@anl.gov.

*Zizhong Chen* received a bachelor's degree in mathematics from Beijing Normal University, a master's degree degree in economics from the Renmin University of China, and a PhD degree in computer science from the University of Tennessee, Knoxville. He is an associate professor of computer science at the University of California, Riverside. His research interests include high-performance computing, parallel and distributed systems, big data analytics, cluster and cloud computing, algorithm-based fault tolerance, power and energy efficient computing, numerical algorithms and software, and large-scale computer simulations. His research has been supported by National Science Foundation, Department of Energy, CMG Reservoir Simulation Foundation, Abu Dhabi National Oil Company, Nvidia, and Microsoft Corporation. He received a CAREER Award from the US National Science Foundation and a Best Paper Award from the International Supercomputing Conference. He is a Senior Member of the IEEE and a Life Member of the ACM. He currently serves as a subject area editor for *Elsevier Parallel Computing* journal and an associate editor for the *IEEE Transactions on Parallel and Distributed Systems*.

*Franck Cappello* is the director of the Joint-Laboratory on Extreme Scale Computing gathering six of the leading high-performance computing institutions in the world: Argonne National Laboratory, National Center for Scientific Applications, Inria, Barcelona Supercomputing Center, Julich Supercomputing Center, and Riken AICS. He is a senior computer scientist at Argonne National Laboratory and an adjunct associate professor in the Department of Computer Science at the University of Illinois at Urbana–Champaign. He is an expert in resilience and fault tolerance for scientific computing and data analytics. Recently he started investigating lossy compression for scientific data sets to respond to the pressing needs of scientist performing large-scale simulations and experiments. His contribution to this domain is one of the best lossy compressors for scientific data set respecting user-set error bounds. He is a member of the editorial board of the *IEEE Transactions on Parallel and Distributed Computing* and of the ACM HPDC and IEEE CCGRID steering committees. He is a fellow of the IEEE. E-mail: cappello@mcs.anl.gov.